\PassOptionsToPackage{dvipsnames,svgnames}{xcolor}
\documentclass[journal]{vgtc}

\onlineid{1916}

\vgtccategory{Research}

\vgtcpapertype{Application/Design Study}

\title{Same Data, Different Audiences: Using Personas to Scope a Supercomputing Job Queue Visualization}

\author{
  \authororcid{Connor Scully-Allison}{},
  \authororcid{Kevin Menear}{},
  \authororcid{Kristin Potter}{},
  \authororcid{Andrew McNutt}{},\\
  \authororcid{Katherine E. Isaacs}{}, and
  \authororcid{Dmitry Duplyakin}{}
}

\authorfooter{

  \item
  	Connor Scully-Allison, Andrew McNutt and Katherine E. Isaacs are with the Scientific Computing and Imaging Institute (SCI) and the University of Utah Khalert School of Computing.
  	E-mail: {cscullyallison|andrew.mcnutt|kisaacs}@sci.utah.edu
  \item
    Kevin Menear, Kristin Potter, and Dmitry Duplyakin are with the National Renewable Energy Laboratory
    Email: {kevin.menear|kristi.potter|dmitry.duplyakin}@nrel.gov
}

\abstract{
Domain-specific visualizations sometimes focus on narrow, albeit important, tasks for one group of users. This focus limits the utility of a visualization to other groups working with the same data. While tasks elicited from other groups can present a design pitfall if not disambiguated, they also present a design opportunity---development of visualizations that support multiple groups. This development choice presents a trade off of broadening the scope but limiting support for the more narrow tasks of any one group, which in some cases can enhance the overall utility of the visualization. We investigate this scenario through a design study where we develop \textit{Guidepost}, a notebook-embedded visualization of supercomputer queue data that helps scientists assess supercomputer queue wait times, machine learning researchers understand prediction accuracy, and system maintainers analyze usage trends. We adapt the use of personas for visualization design from existing literature in the HCI and software engineering domains and apply them in categorizing tasks based on their uniqueness across the stakeholder personas. Under this model, tasks shared between all groups should be supported by interactive visualizations and tasks unique to each group can be deferred to scripting with notebook-embedded visualization design. We evaluate our visualization with nine expert analysts organized into two groups: a "research analyst" group that uses supercomputer queue data in their research (representing the Machine Learning researchers and Jobs Data Analyst personas) and a "supercomputer user" group that uses this data conditionally (representing the HPC User persona). We find that our visualization serves our three stakeholder groups by enabling users to successfully execute shared tasks with point-and-click interaction while facilitating case-specific programmatic analysis workflows.
}


\usepackage{graphicx}

\teaser{
  \centering
  \includegraphics[width=\linewidth, alt={}]{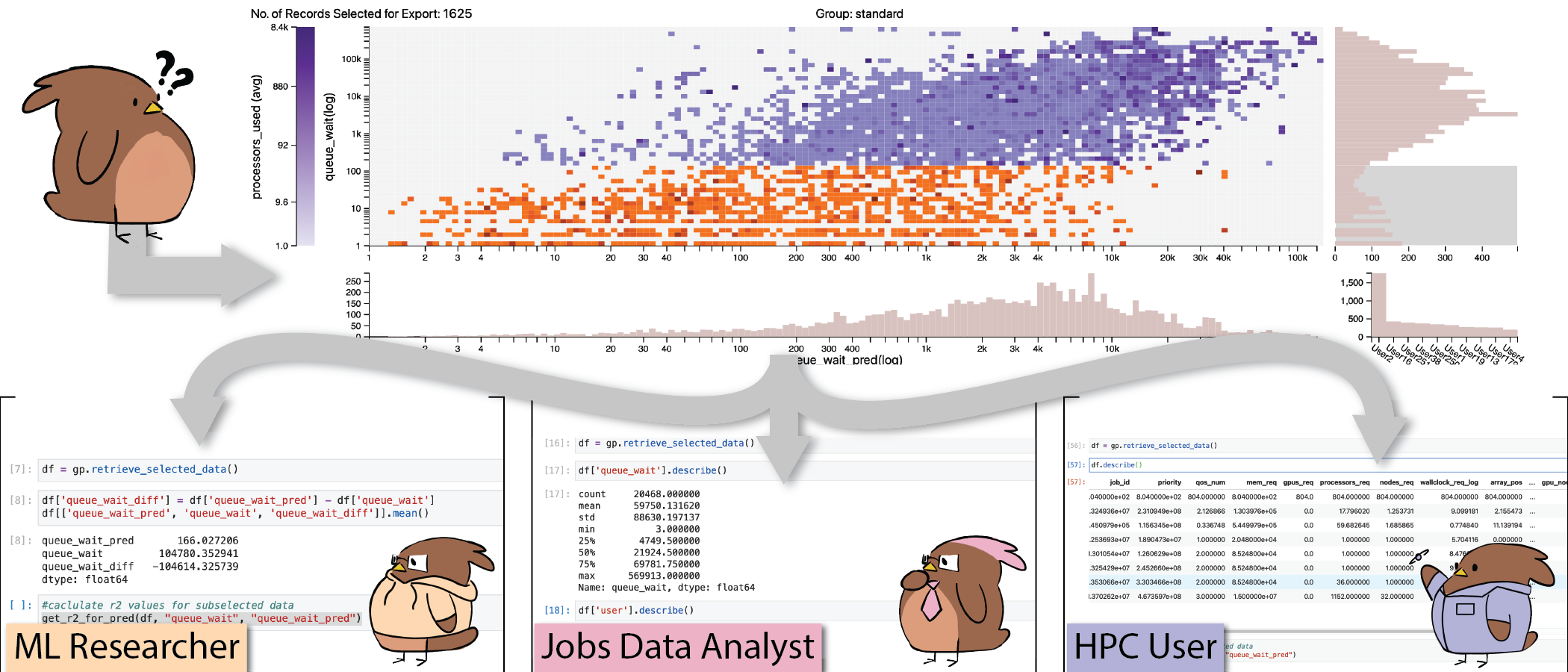}
  \caption{Many different types of users care about supercomputing data, but not all of them have the same needs. \texttt{Guidepost} seeks to support varied users interacting with this type of data. After loading data in a Jupyter notebook and getting their bearings using \texttt{Guidepost}, our individual analysts export interesting selections of data back into the Python code context of the notebook. They then ``flock'' to their own workflows, having used the visualization as a ``guidepost'' directing their analytical migrations from data to insight.}
  \label{fig:teaser}
}

\graphicspath{{figs/}{figures/}{pictures/}{images/}{./}}

\usepackage{tabu}
\usepackage{booktabs}
\usepackage{lipsum}
\usepackage{mwe}

\usepackage{mathptmx}
\usepackage{pifont}
\usepackage{tabularx}

\newcommand{\hu}{\colorbox{blue!30}{\footnotesize \textsc{HPC User}}}
\newcommand{\jda}{\colorbox{purple!30}{\footnotesize \textsc{Jobs Data Analyst}}}
\newcommand{\ml}{\colorbox{orange!30}{\footnotesize \textsc{ML Researcher}}}
\newcommand{\hus}{\colorbox{blue!30}{\small \textsc{HPC User}}}
\newcommand{\jdas}{\colorbox{purple!30}{\small \textsc{Jobs Data Analyst}}}
\newcommand{\mls}{\colorbox{orange!30}{\small \textsc{ML Researcher}}}

\newcommand{\sharedTask}[1]{ST$_{\textrm{#1}}$}

\newcommand{\partialTask}[1]{PT$_{\textrm{#1}}$}

\newcommand{\dTask}[1]{DT$_{\textrm{#1}}$}

\newcommand{\guidepost}{\texttt{Guidepost}}

\newcommand{\inlinesec}[1]{\vspace{0.5ex}\noindent\textbf{#1}}

\begin{document}

\renewcommand{\figureautorefname}{Fig.}
\renewcommand{\sectionautorefname}{Sec.}
\renewcommand{\subsectionautorefname}{Sec.}

\firstsection{Introduction}

\maketitle

Custom-built domain-specific visualizations sometimes focus on narrow, albeit important, tasks
by targeting one group of users.
This confined focus can limit the utility to other groups with interest in the same data.

We suggest that careful consideration of these multifaceted groups---such as by better defining user groups and tracking the origins, importance, and scope of tasks with respect to them---can yield visualizations more useful to all involved parties.

To this end, we sketch a {\em persona}-focused\cite{cooper_about_2003, cooper_inmates_2004} design process

that serves multiple groups with partially divergent tasks.

In this approach, we use personas as surrogates for analyst groups to organize elicited tasks and identify reasonable design goals based on overlapping needs between personas.
By explicating the needs that various groups have, we can form a clear picture where a single high-level visualization intervention will provide the greatest utility for the most users, and then provide explicit off-ramps from that interface so that persona specific tasks can be supported.
We then reserve bespoke visualization development efforts for tasks performed by multiple groups. For needs belonging to a single group, we provide interlocked tools that allow individual analysts to fall back into their standard workflows.

We investigate this approach in a visualization design study~\cite{sedlmair_design_2012} with a group of experts at a  at US-based Federally Funded Research and Development Center (FFRDC) who are investigating various aspects of High Performance Computing (HPC) data---specifically, how programs (``jobs'') are scheduled to run on supercomputers.

We identify three distinct groups of users which we cast into the following personas: \hu, \jda, and \ml. We collected tasks from each group (in both informal and formal elicitation meetings) and then selected which tasks our visualization would directly serve based on their usage across groups, relegating more unique tasks to scripting.

From this analysis, we developed \guidepost{}, an interactive visualization embedded in Jupyter notebooks. \guidepost{} uses a collection of pez plots~\cite{walsh_pez_2023} to provide a configurable overview of HPC jobs data. The ease of flexibly configuring which attributes dominate the layout supports the goals of the different personas. \guidepost{} also provides interfaces for visualizing multiple supercomputer job queues, changing attributes under investigation, and filtering by categorical variables.
To support tasks specific to individual groups, we enable fluid context-switching from our visualization to the notebook's scripting interface.
We do so by allowing users to select subsets of data in the visualization and then retrieve them~\cite{manz_anywidget_2024}  as pandas~\cite{the_pandas_development_team_pandas-devpandas_2020} dataframes in code.

We validate this design with nine experts from three FFRDCs who analyze HPC jobs data. We find that this design conforms well to user needs (by their description) and that they were largely successful with our tasks, employing diverse strategies to analyze the data. Structuring visualization interventions in such a way that they allow their users to ``fall back'' into a workflow they are already familiar with provides benefits to users and developers. This approach supports users' desires for the expressiveness that coding-based workflows provide and reduces the development effort required to support each user's tasks explicitly.

In summary, the primary contributions of this work are:
\begin{itemize}
    \itemsep -0.5em
    \item An example of a design process for synthesizing tasks of multiple user groups (\autoref{sec:design}).

    \item A design for a notebook-embedded overview visualization for supercomputer jobs data (\autoref{sec:guidepost})

\end{itemize}

A video of \guidepost{} as well as a demonstration notebook is available in our supplementary materials.

Bespoke user centered visualizations can serve some people all of the time at the cost of serving some of the people none of the time. Whereas general purpose visualizations serve all of the people some of the time at the cost of not serving certain needs all of the time. We demonstrate that carefully placed interventions and not trying to serve every case can be a valuable design strategy for building sustainable user-centered visualizations that support multiple populations working with the same data.

\section{Related Work and Background}

    We present related work on personas in visualization design and domain-specific visualization solutions for HPC jobs data. We also provide background on the domain of High Performance Computing program schedulers and program management.

    \subsection{Personas in Visualization Design}

    Introduced by Cooper\cite{cooper_inmates_2004}, a persona is a ``fictional person'' that models a potential or current user of a software tool \cite{salminen_literature_2020}. By contrast to task-based approaches, a persona provides designers of a system with a qualitative representation of a potential user. Personas are often described narratively and characterized by their needs and goals relative to the system being designed \cite{miaskiewicz_personas_2011}.

    Despite their common adoption in user-centered computer science subdomains like software engineering\cite{cooper_about_2014, grudin_personas_2002, pruitt_persona_2010} and Human Computer Interaction\cite{acuna_hci_2012, goodwin_getting_2018}, personas are relatively under-explored as a method for describing user needs in the area of data visualization. Some recent works\cite{van_den_brandt_understanding_2025, li_knowledge_2024} have adopted the use of personas as descriptive models to summarize users of classes of visualization tools. Outside of this, few design studies explicitly adopt or use personas as a tool for characterizing and understanding user needs.

    Other works in visualization hint at the idea of personas without using the term directly. Burns et al. explore the idea of an oft-referenced pseudo-persona used in visualization design: the visualization ``novice'' \cite{burns_who_2023}. Sedlmair et al.'s~\cite{sedlmair_design_2012} design study methodology evokes the idea of personas through project "roles" to describe groups of individuals archetypically. Many subsequent works use these roles to define their end users as \textit{front-line analysts}~\cite{devkota_ccnav_2021, elshehaly_qualdash_2021}, however this ``role'' is too broad to describe user needs at the specific level required by our project which aims to be very deliberate about designing for many user groups who could all be called \textit{front-line analysts}.

    In visualization research, user needs are more commonly described as a discrete set of tasks that a potential visualization should support \cite{shneiderman_eyes_1996, munzner_nested_2009, munzner_visualization_2014}.
    This approach provides a clear mapping between specific user needs and what a visualization designer builds. When executed well, this approach can make implementation and validation of a visualization system very straightforward, but also runs the risk of over-formalizing otherwise abstract and malleable user needs.

    By contrast, personas offer many unique benefits compared to task-based requirements specification models.
    They are linked with designers focusing more on target users' needs \cite{miaskiewicz_personas_2011, grudin_personas_2002, long_real_2018}. Closer to our approach to developing for multiple user groups is prior work in the product design that identifies personas as beneficial in ``narrowing down the users being designed for'' \cite{miaskiewicz_personas_2011, pruitt_persona_2010}. We use this aspect of personas to make fine distinctions between similar user groups.

    \subsection{Jobs, Schedulers, Queue Time Prediction}

    When a program is submitted for execution on a supercomputer, additional parameters such as program inputs and resource needs (e.g., number of processors) must also be sent. This bundle of metadata specifies the ``job'' a user is asking the supercomputer to perform. This job request is placed into a {\em queue} until it is scheduled by a special type of refereeing software called a ``scheduler''~\cite{ahn_flux_2020, yoo_slurm_2003}. There are numerous concerns that need to be taken into account when scheduling jobs, like the availability of the requested resources or the priority of the job. The scheduling software helps to manage these variables in a fair way.

    Schedulers enable system administrators to organize the resources that make up a supercomputer---called ``nodes''---into logical ``queues'' or ``partitions.'' These partitions can sometimes require that certain hardware is used for a job (e.g., GPU code needs to go to GPU nodes), but otherwise often suggest priorities of when software should run or enforce policies of software behavior. For example, jobs submitted to the ``short'' queue will timeout if they run longer than a set time but may get priority in scheduling to fill holes.

    In addition to providing administrators and users with tools to get their programs to the right computational resources, scheduling software monitors the state of submitted jobs. This data can be logged and stored by the administrators of the HPC system. When analyzed post-hoc, this data describes the behavior of HPC users and systems. Some metrics associated with these jobs are how long they waited on the queue (\texttt{queue\_time}), how long they ran for (\texttt{hours\_used}), who submitted them (\texttt{user}), what queue was it submitted to (\texttt{queue}), and how many resources were requested (\texttt{nodes\_requested}).

    Recently there has been a push by researchers to apply machine learning to this data. Most notably, models have been trained that use job characteristics to predict runtime and/or queue time for jobs that have not yet been submitted to the scheduler system~\cite{menear_tandem_2024, okafor_queue_2024}. Queue time prediction data is of interest to one of our personas---ML engineers working on prediction algorithms.

    \subsection{Visualization Solutions for HPC Jobs Data}

        Visualizing data collected from HPC systems poses unique challenges due to the complexity and scale\cite{sakin_traveler_2023} of the data\cite{bourassa_operational_2019} in addition to the complexity of tasks\cite{devkota_ccnav_2021}.
        These challenges are exacerbated when combined with data produced by machine learning engineers \cite{lepenioti_prescriptive_2020, soltanpoor_prescriptive_2016}, a key stakeholder group in this domain. The study of this data is called HPC Operational Data Analytics (ODA). Netti et al.~\cite{netti_conceptual_2021} highlight that there are many subdomains in this field that necessitate unique visualization solutions.

        Visualization strategies observed in the ODA literature can be broadly categorized into two groups: (1) general purpose visualizations adopted by ODA professionals and (2) domain experts developing ad hoc visualizations. Commonly used software like Open XDMoD~\cite{palmer_open_2015} or Grafana~\cite{labs_grafana_2024} exist in the former camp. These tools support high-level analysis of streaming data but do not enable users to drill down to more detailed low-level analytical tasks. Menear et al.~\cite{menear_mastering_2023} note pitfalls of their approach, such as a lack of consistency in presenting results and the obfuscation of vital information.

        Among domain-driven solutions that visualize HPC jobs data, notable examples are HiperJobVis\cite{nguyen_hiperjobviz_2019} and JobViewer\cite{dang_jobviewer_2022}. HiperJobVis focuses primarily on HPC users and how their behavior affects supercomputers. While useful for system administrators, it offers less utility when trying to examine other factors that affect system performance. By contrast, our solution supports user data but focuses instead on queue-wide data. JobViewer focuses on high-level analysis of HPC systems in a purely point-and-click context with specialized charts. Our work aims to be more specific to the analysis of jobs data and support the flexibility afforded by code-based analyses in notebooks.

\section{Visualization Design Process}
    \label{sec:design}
\begin{figure}
        \centering
        \includegraphics[width=\linewidth]{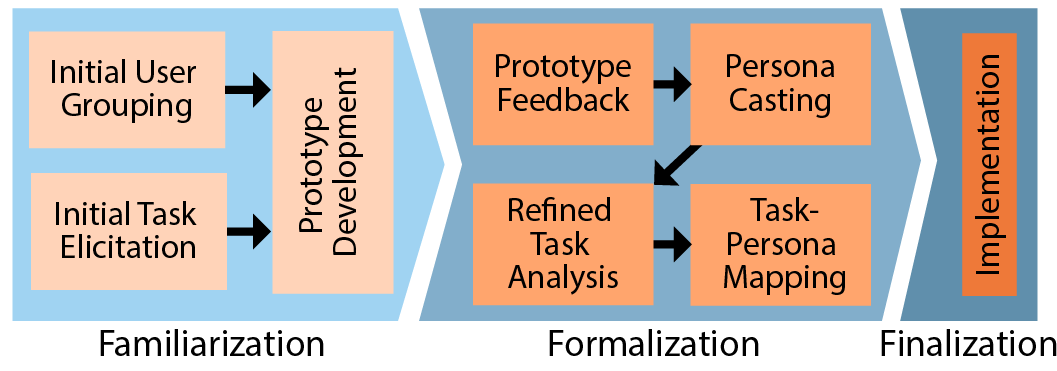}

        \caption{The phases of the design process used to develop \guidepost.
        In the familiarization phase, we engaged in task elicitation and user grouping in parallel, with the goal of developing a prototype that could serve more than one user group's needs.
        Next, in formalization, we used the feedback obtained from interviews centered around the prototype to refine personas, tasks, and make a formal mapping between which personas requested which tasks.
        Finally, in finalization, we reify this model as an implemented system.}

        \label{fig:design-process}
        \vspace{-2em}
    \end{figure}

    We present our design process for developing \guidepost{}. We have retroactively organized our process to make it more easily followed for future projects (\autoref{fig:design-process}). We retroactively organize this design process into three phases: familiarization (during which we get a lay of the land), formalization (where we refine that map), and finalization (where we develop the final tool this process was designed to build).

    \subsection{Familiarization}

    To begin, we sought to get acquainted with the users and other stakeholders for the domain. This phase had two (parallel) stages: modeling the personas and modeling the tasks.
    This culminated in a usable prototype that accurately displayed user data.
    We used this prototype to gather feedback on our design for the next phase.

    \inlinesec{Initial User Grouping:} At the start of this project, we initially cast individuals into Design Study Methodology~\cite{sedlmair_design_2012} roles: \textit{gatekeeper}, \textit{front-line analyst}, and \textit{fellow tool builder}.
    As the project progressed, we found that individuals had more nuanced (and less homogeneous) roles than we initially thought.
    To wit,  individuals we spoke with were often \textit{front-line analysts} in addition to their other roles, supplying important tasks that used the same data but varied in their analysis goals.
    While each of these could be cast as front-line analysts, the differences in tasks made it hard to see them as carrying out the same role.
    For example, some potential users were interested in reducing queue times, while others focused on understanding how to make more accurate queue time predictions.
    We made informal attempts to subdivide these potential users into the following ``groups'' that resembled personas: \texttt{Operational Data Analyst} and \texttt{HPC User}.

    We used these groups to track the origin of tasks and identify commonly mentioned tasks. Since this was done in parallel with task elicitation, we also used obtained tasks to identify commonalities and differences between groups.

    \inlinesec{Inital Task Elicitation:}  Our task elicitation began with two potential users from the \texttt{Operational Data Analyst} group, with regular weekly meetings. The first author discussed needs and current workflows with these users.  These elicitations were supplemented with a more structured meeting with a single individual who fit into the \texttt{HPC User} group.
    Across these meetings, the lead author identified the following low-level tasks that describe how individuals might be exploring this data:

        (1) Understand the uncertainty of queue time predictions.
        (2) Identify likely upper and lower bounds of queue time predictions.
        (3) Compare queue time predictions across queues.
        (4) View prediction(s) in the context of historical data.
        (5) Identify potential opportunities to reduce queue time.
        (6) Explore machine learning model accuracy.
        (7) Understand how variables may relate to prediction accuracy.
        (8) Summarize model accuracy.

    \inlinesec{Prototype:}  \label{sec:prototype}

    Using this initial modeling we developed several prototypes, which focused on the \texttt{Operational Data Analyst} group, featuring detailed plots of attributes versus runtime predictions. After multiple iterations and discussions with stakeholders, we shifted our focus to analysis of prediction accuracy.
    We designed a streamlined approach that plotted predicted queue time against actual historical queue times. See supplement for a full description and figures.

    This prototype focused on analyzing queue time prediction data produced by a model developed by one of the \texttt{Operational Data Analysts} we regularly met with and formed the basis of what eventually became \guidepost{}.

However, during this phase elements deemed superfluous were dropped. Most notably, Quantile Dot Plots \cite{kay_when_2016} were identified by our users as unnecessary to this expert population.
Explanation of the encoding rules for the QDP was sometimes difficult for our users and resulted in requests for a more standard probability density function visualization. Our user base comprises groups largely familiar with probability and statistics.
In cases like ours, a more traditional uncertainty visualization may serve the same purpose as a visualization that is more effective at explaining uncertainty to a general populace.

\begin{table*}[t]
    \centering
    \footnotesize
    \setlength{\tabcolsep}{4pt}
    \caption{Task descriptions.
    Shared tasks (ST) are performed by all personas, Partially-shared (PT) by two, and Disjoint (DT) by only one persona.
    }
    \begin{tabularx}{7.25in}{|p{2.1cm}|p{3.2cm}|X|}
        \hline
        \textbf{Task} & \textbf{Name} & \textbf{Persona-Oriented Descriptions} \\
        \hline
  \hline
    \sharedTask{compare\_q} & Compare queue time across queues & All personas are interested in comparing queue time across queues. The goals of their analysis differ with \hus{} wanting to identify the optimal queue to submit to, and \jdas{} and \mls{} seeking to see broader patterns within and between each queue. \\
    \hline
    \sharedTask{assess\_q\_wait} & Understand how queue time is influenced by other variables on the dataset & \hus{} wants to understand how they can change their job configurations to reduce queue times. \mls{} is interested in what variables might be good features to select for in an ML model. \jdas{} is interested in seeing whether the behavior of the system is consistent with expectations. \\
    \hline
    \sharedTask{relate\_jobs} & Relate jobs data to individual users or user groups & \hus{} wants to view their own jobs data (or the jobs data of their research group) and assess if they need to change usage patterns. \jdas{} wants to identify power users and understand their impact on the system, especially with investigations like failure analysis \cite{lu_failure_2013}. \mls{} wants to assess how a particular user's data may be impacting their model. \\
    \hline\hline
    \partialTask{inspect\_jobs} & Inspect jobs data over time & \hus{} wants a temporally organized record of jobs data for project planning. If they set milestones that avoid historically busy times, this can reduce queue competition and wait times. For \jdas{}, viewing data over time is the default perspective they want on the data. Much of their expectations and analyses hypotheses are informed by seasonal variance in system usage. \\
    \hline
    \partialTask{filter\_jobs} & Filter using general categorical features & \jdas{} wants to slice the data and filter using many different categories that group jobs logically: the priority of a job, the exit codes for error analysis, and the day of the week that a job ran. These features and more can be important to understanding patterns of system behavior. The \mls{} persona is interested in seeing the impact of a particular categorical feature on their model's behavior. Often used to assess if a particular feature should be included, removed or re-engineered in some way. \\
    \hline
    \partialTask{data\_context} & Assess queue-time predictions in the context of historical data & \mls{} is interested in viewing predictions in the context of real data to assess the accuracy of their model from a high-level perspective. This is primarily an exploratory task for \mls{} because they do not know exactly where accuracy problems will be found or what they want to do when identified. \jdas{} wishes to understand if a deployed model that makes queue time predictions for users is working properly. \\
    \hline\hline
    \dTask{time\_compare} & Compare trade-off between queue time and run time & \hus{} wants to make specific assessments about whether they can reduce their queue times at the cost of increased run times by allocating fewer resources to a job. \\
    \hline
    \dTask{view\_bounds} & Estimate likely upper bounds of queue time predictions & \hus{} wants to understand how long they can realistically expect to wait for a planned job to run. They use both historical data and predictive models to accomplish this. \\
    \hline
    \dTask{explain\_system} & Communicate state of HPC system to stakeholders & Many stakeholders are interested in the behavior of HPC systems. They want metrics and analytics that show how things are running and what may need to change to hit organization-level targets. \jdas{} is primarily responsible for getting this information into the hands of these stakeholders. \\
    \hline
    \dTask{explore\_model} & Explore machine learning model predictions & \mls{} wants to perform exploratory analysis on their model's predictions. They want to perform statistical assessments on how variables are co-related with one another and weigh their impact on model accuracy. \mls{} also wants to use statistical methods to reduce model accuracy to summaries that describe a model in quantifiable terms. \\
    \hline
    \dTask{uncertainty} & Understand the uncertainty of a queue time prediction & \mls{} seeks to understand how their model's predictions are spread across a range of potential outputs and how to narrow that range to an accurate representation of the training data. This is categorized as a distinct task from \dTask{view\_bounds} because the focus in evaluating the uncertainty is much wider for \mls{}. \\
    \hline
        \end{tabularx}
    \label{tab:task-table}
    \vspace{-2em}
\end{table*}

    \begin{figure*}
        \centering
        \includegraphics[width=.85\linewidth]{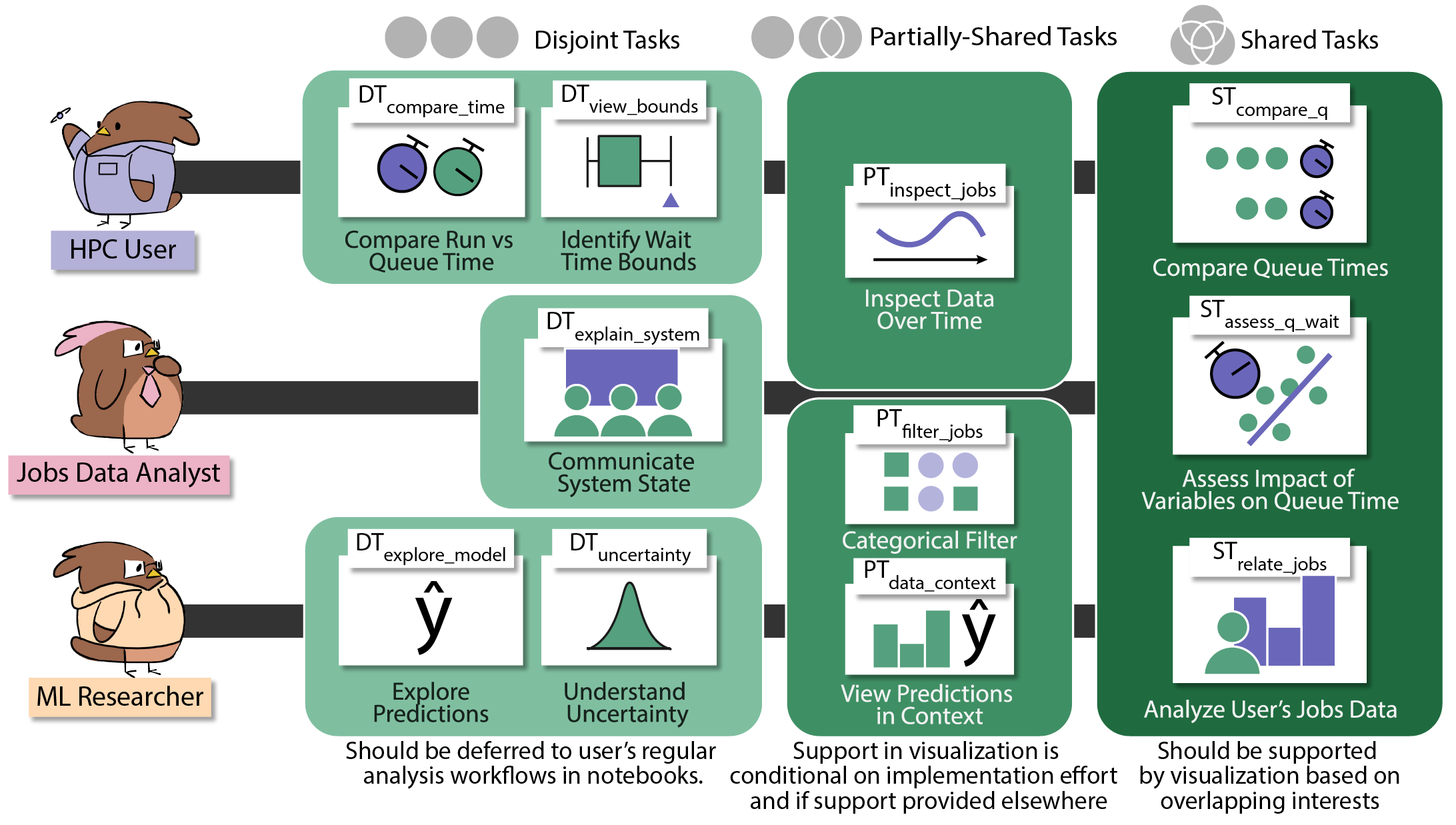}
        \caption{The results of the task analysis we executed for \guidepost. Each group is represented by a persona to the left. Tasks (rounded rectangles) overlap the black `tracks' of the personas who perform them, with tasks on the left belonging to single personas and those on the right belonging to all personas.}
        \label{fig:tasks}
        \vspace{-2em}
    \end{figure*}

 \subsection{Formalization}
     \label{sec:personas}

Next, we sought to formalize what we had learned about our users and their domain. To do so, we first sought feedback on our prototype and used that feedback to

to formalize that model of persona and task, and then create a map between them.

    \inlinesec{Prototype Feedback:} We conducted (N=8) semi-structured interviews with potential users of our tool, namely experts working with HPC systems or analyzing HPC data. We acquired this pool of users through direct solicitation at the FFRDC where this study took place.
    Interviews involved showing users the first prototype visualization, explaining each of its component views, and asking for feedback.

    Beyond design feedback, we asked users to describe the way they might use HPC jobs data as part of their workflows and what questions they have about the operation of the system as a means of refining our task and persona models. We analyzed this data using an open coding process followed by a high-level grouping of similar ideas into categories~\cite{glaser_discovery_1968} and characterized personas based on the overlapping goals and interests different participants exhibited.

    \inlinesec{Persona Casting}     Following guidance from industry resources\cite{laubheimer_3_nodate}, we used the data solicited from our feedback sessions to refine our user groups into personas.

    From our initial two groupings, we developed the three personas described below. While the \texttt{HPC User} grouping was mostly unchanged, the \texttt{Operational Data Analyst} grouping was broken into two as we observed distinct tasks and goals that suggested two groups. Primarily, some of our \texttt{Operational Data Analysts} showed interest in seeing descriptive data, which the prototype was not focused on, while others were inclined towards the predictive data our prototype did focus on.
        We describe each of the personas:

    \newcommand{\question}[1]{``\emph{#1}''}

    \noindent\hu: They are a domain scientist (physical scientist, life scientist, energy researcher, etc.) that runs simulations on supercomputers. This individual's interactions with a supercomputer system are often managed by custom scripts, executed in a remote terminal that configures run-time parameters for their simulations. The \hu's interaction with jobs data is often limited to viewing their own data and that of their team to assess historical usage, job conditions, and plan for future allocation requests. When looking at more data of the system than just their own, their goals are to understand what configurations or behavior they can change to reduce or decrease their queue times and software run times. Examples of questions they ask are: \question{When is the supercomputer used the least?} or \question{Could I get a faster queue time if I chose a different queue?}

    \noindent\jda: They are a researcher or system administrator that focuses on studying jobs data with broad exploratory questions about the behavior of an HPC system. They may use the supercomputer to perform system tests or benchmark software; however, they mostly focus on understanding how a supercomputer system works. They want to understand how user behavior can impact a supercomputer and identify the source of issues that cause software or hardware crashes or unexpected lag on a system. They want a broad view of the data that shows patterns and correlates many variables that describe supercomputer ``jobs.'' Examples of analysis questions they ask are: \question{Did a particular program or user cause this crash or slowdown on the system?} or \question{Could the job scheduler be working better?}

    \noindent\ml: They are a specialized researcher exploring how to use machine learning to understand HPC systems better. They use historical jobs data to build machine learning models that predict future behavior about a supercomputer system. They want to make the \hu{}'s life easier by aiding their decision-making process. They work closely with the \jda{} to acquire and validate jobs data. \ml is mostly interested in the accuracy of their models; however, they also want to understand which features describing supercomputer jobs are most descriptive of historical trends for feature engineering\cite{murel_what_2024}. Questions they may ask are: \question{Are there any outliers in my set impacting prediction accuracy?} or \question{Are job runtime and queue time correlated?}

\inlinesec{Refined Task Analysis}

    Using interview transcripts collected from the prototype feedback interviews and notes from weekly meetings with the stakeholders we worked most closely with, we refined and expanded our task analysis. We used open coding~\cite{glaser_discovery_1968} to identify new tasks mentioned by users that were not captured by our prototype.
    We used this opportunity to speak with more stakeholders to modify our existing tasks to more accurately fit into one persona's needs or another.

    In~\autoref{tab:task-table}, we show our analysis of our elicited tasks organized by their commonality between identified groups of users.

\inlinesec{Task-Persona Mapping}

    Finally, we creating a mapping between our tasks and the personas who would need to perform them. We summarize this mapping in \autoref{fig:tasks}.
    Despite the apparent similarities of some of the roles, each persona has a set of distinct tasks that are unique to them. For instance, only the \mls{} cares about exploring predictions made over the data. In contrast, there are some tasks that all personas care about, such as comparing queue times.
    This highlights the inherent wickedness of trying to navigate a multi-stakeholder domain: some solutions will support some users partially, while others will support some users totally.
    Navigating this challenge is down to the specific values and strategies associated with the work.

    \subsection{Finalization}

Finally, we put this model into practice by designing a system around it.
In the next section (\autoref{sec:guidepost}), we describe in detail the design of the visualization built, \guidepost{}.

Here, we make note of our strategies for parsing our task-persona mapping.
An important insight in this design is that many of these tasks can be performed, with difficulty, using a typical analysis workflow in a computational notebook.
Observing this, we partition tasks into those that should continue to use that workflow (because they are rare relative to the larger group) and those that would benefit from an interactive visualization tool.
In an unlimited resource environment, dedicated solutions could be built for each of these personas, but given the fixed resources associated with this (or any) design study, we tried to select our partition in a manner that maximized our impact for the most users.

In our partitioning, we
prioritized implementation of tasks that were fully overlapping among our personas, again, such as comparing queue times.
In contrast,

fully disjunct tasks, expressed by only one user group, were deferred to scripting to maintain the accessibility of the visualization to all user groups.

    We evaluated partially overlapping tasks on a case-by-case basis.
    We considered the consequences of implementing a feature for a particular task from many angles.
    A significant factor was how support might complicate the visualization for users who are not performing this task.
    Another was evaluating the benefit to the user against the difficulty of implementation.
    If support already exists for a task in a library (like pandas) or a task could be supported by another visualization software that could be used without substantial difficulty, we deemed it not worth implementing. For example, many domain-specific machine learning libraries \cite{waskom_seaborn_2021, virtanen_scipy_2020, team_theano_2016} provide myriad tools for exploring machine learning predictions.

    This approach limits the potentially wasteful expenditure of resources on development for an increasingly small audience.

    While this specific process was guided by our own experience conducting this research, we suggest that explicit consideration of the range of tasks, the personas who perform those tasks, and strategies to guide implementations to maximize impact among those personas is worthwhile for any design process with multiple user types.

\begin{figure*}
    \centering
    \includegraphics[width=.85\linewidth]{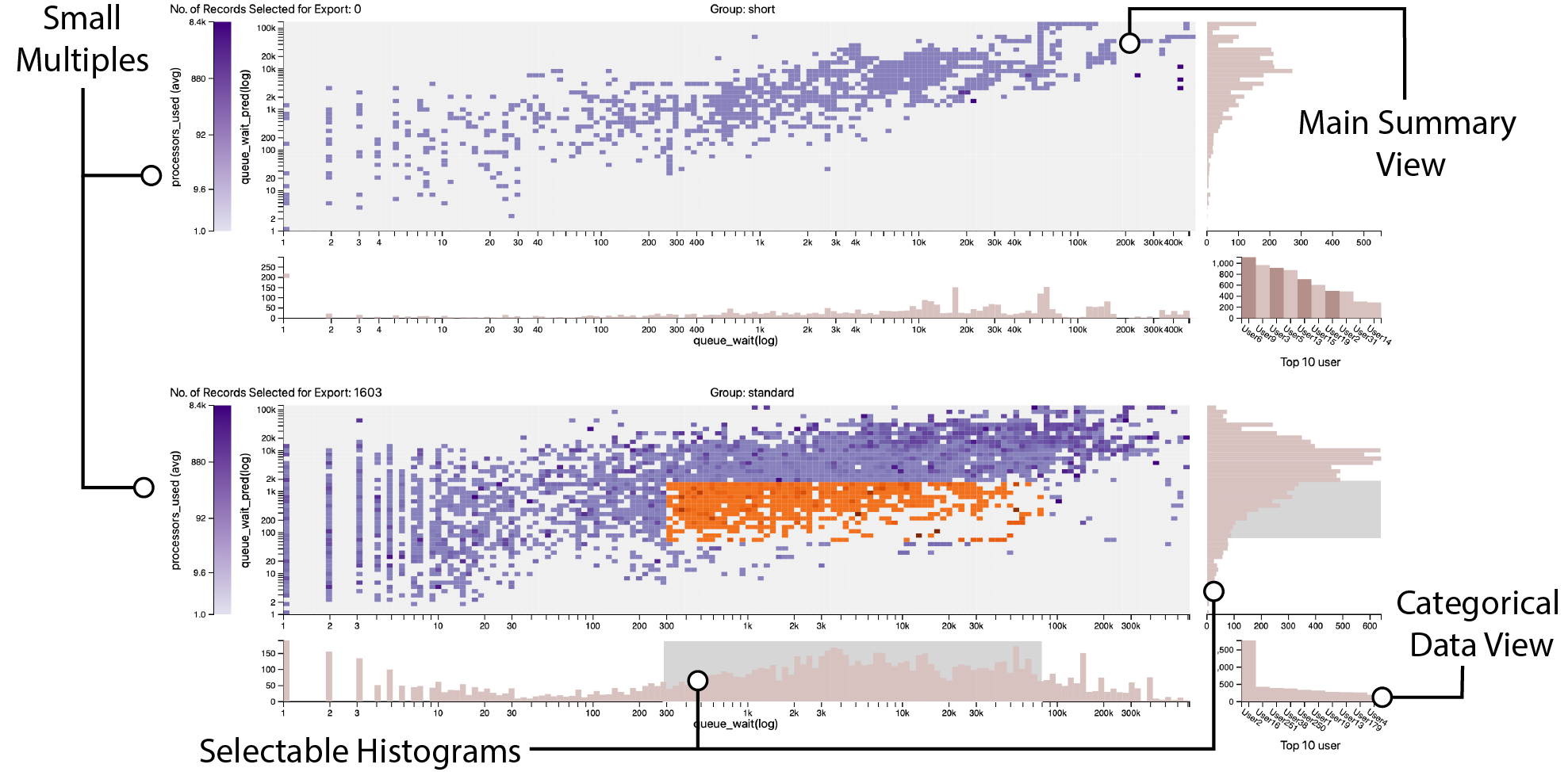}
     \caption{Two instances of the \guidepost{} visualization as small multiples to represent two supercomputer queues. \guidepost{} comprises four linked views: the \textit{Main summary View}, the \textit{Categorical Data View}, and the \textit{Selectable Histograms}.}
    \label{fig:full_system}
    \vspace{-2em}
\end{figure*}

\section{Guidepost}
    \label{sec:guidepost}

    Using the personas, task analysis, and task-persona mappings from (Sec.~\ref{sec:design}), we developed \guidepost{}: an interactive, filterable, overview visualization designed for notebook-based workflows.
    This interactive visualization gives users an overview of their data and supports widely used exploration features that seamlessly connect with a notebook-based analysis workflow. \guidepost{}'s primary view is a summary visualization that shows multiple features of the data by x and y position and color. This view is supplemented with two histograms showing distributions of the data as well as a bar chart which filters the data on click.

\subsection{Visualization Configuration}

    Leveraging \guidepost's notebook-embedded workflow, we opted to defer the configuration of various visual encodings to a Python \texttt{dict}. Our rationale was that we wanted to encourage users to return to the scripting side of the notebook frequently enough to consider the visualization in the wider context of the notebook. Further, we saw an opportunity to leverage the scripting interface to provide customization functionality to our users without cluttering the visualization with point-and-click selection interfaces.

    % % \begin{verbatim}
    % \begin{minted}{python}
    % gp.vis_configs = {
    %     'x-axis': 'submit_time',
    %     'y-axis': 'queue_wait',
    %     'color': 'nodes_requested',
    %     'categorical': 'user',
    %     'facet_by': 'partition'
    % }
    % \end{minted}
    % % \end{verbatim}

    For the rest of this section we will discuss features of the visualization in the context of these specific configurations.

\subsection{Small Multiples}
     Users can create small multiples of \guidepost{} using any repeating categorical variable that groups the data. By default, our interface creates small multiples that facet by queue to support our user tasks. We identified that all user groups were interested in comparing across queues (\sharedTask{compare\_q}). We also observed from our machine learning-oriented prototype that one of the variables most strongly correlated with queue time (\sharedTask{relate\_jobs}) was the queue to which a job was submitted. Finally, we observed that our user base strongly considered queues to be semantically distinct categories on the dataset and often grouped by queue in their analysis.

     With ``queue'' as the dominant facet, our unit design shows data associated with a single queue. Within each unit visualization we have a \textit{Main Summary View}~(\ref{sec:PezPlot}), two \textit{Querying Histograms}~(\ref{sec:Histograms}), and a \textit{Categorical Data View}~(\ref{sec:CatDataView}). Each \guidepost{} unit also has a legend that describes the current number of selected records and a color map showing the extents of the ``color'' variable (e.g. `nodes\_requested') mapped to the \textit{Temporal Summary View}.

\subsection{Main Summary View}
\label{sec:PezPlot}

    Within each \guidepost{} unit, the central view of the data comes is the \textit{Main Summary View}: a grid of rectangles framed by two histograms. The horizontal axis encodes a \texttt{datetime}, \texttt{integer}, or \texttt{float} variable associated with each job (\partialTask{data\_context}). In \autoref{fig:full_system}, this variable is a measured queue time associated with each job. The vertical axis encodes any other user-specified numerical attribute. As the majority of our data is heavily skewed, we set it to log scale by default. In \autoref{fig:full_system}, this axis is configured to show "queue\_wait". This attribute set is the default encoding to support \sharedTask{compare\_q}, \sharedTask{assess\_q\_wait}, and \dTask{explain\_system}.

    Data is binned first into columns according to the data type and scale of the x-axis variable. A linear scale is used for binning by default. For \texttt{integers} and \texttt{floats}, variables with a range divided by two orders of magnitude are binned based on log scales. Since log-scaled, numerically even intervals would not produce visually even bins, we define a function that calculates ranges for a fixed number of intervals evenly spaced on a log scale. This process is similar to NumPy's "logspace" algorithm~\cite{numpy_numpylogspace_2024}. For \texttt{datetime} variables, values are binned into hours, days, or months depending on the total range.

    Within each column, we divide the data into row-oriented bins. The y-axis variable cannot be a \texttt{datetime} but otherwise behaves in the same fashion as the x-axis. The color variable (e.g. ``nodes\_requested'') dictates how each rectangle is colored. The darkness of the color is produced by an aggregation over all the data points binned into each rectangle. This ``color'' variable can be changed to show any numerical data we have on our dataset (\partialTask{inspect\_jobs}) and may want to correlate with the variables depicted on the x and y axes.

\subsection{Querying Histograms}
\label{sec:Histograms}
    The \textit{Main Summary View} is supported by two histograms that show distributions of job records. The x-axis-aligned histogram shows the number of jobs associated with each column of the \textit{Main Summary View}. When a \texttt{datetime} is encoded on the x-axis, this choice enables users to quickly identify times where a supercomputer is used more or less frequently (\partialTask{data\_context}). It also provides a broad view of events that may affect system usage. For example, periods of system downtime are empty of jobs.

    The second histogram is vertically oriented along the right-hand side of our main plot and corresponds to the y-axis variable (e.g. "queue\_wait"). It shows the number of records in each row along the y-axis and gives a quick indication of the distribution of our y variable. When plotting "queue\_wait" on this axis, this distribution shows how long the majority of jobs wait to run. Rapid comparisons between queues  (\sharedTask{compare\_q}) are enabled by this view.

\subsection{Categorical Data View}
\label{sec:CatDataView}

The final view in each \guidepost{} unit is a bar chart nestled in the lower right. The bar chart encodes up to ten of the most frequently occurring elements of a user-defined categorical variable, such as day-of-week or user. We chose ten as a round number that fits well visually in the space. The height of each bar corresponds to the number of records associated with each category. The categories on the x-axis are sorted by number of records. We default to showing users (as can be seen in \autoref{fig:full_system}). This category, in particular, was often referenced in stakeholder interviews and highlighted in related work~\cite{nguyen_hiperjobviz_2019}.

\subsection{Interactions and Exporting}
\label{sec:interactions}

Both histograms support brushing interaction that triggers the dynamic exporting capabilities of \guidepost{}, supporting our single-persona tasks by easing access to filtered subsets of the data. Brushing over a histogram also highlights selected cells using an orange color map and updates the "\# of Records Selected: " legend text.

Records selected by brushing are managed by a variable bound between the JavaScript running the visualization and the Python side of the notebook. The records stored in this variable are retrievable in the notebook via a class method on the Python \guidepost{} object: \texttt{retrieve\_selected\_records()}. This method concatenates selected jobs into a single pandas DataFrame organized like the input data. Subsequent re-selections of data automatically update the bound variable with new data. This is the primary method by which we support the tasks specific to one persona: \dTask{time\_compare}-\dTask{uncertainty}. Instead of adding features to \guidepost{} itself, this feature enables quick extraction of ``interesting" subsets of data---as subjectively defined by each persona---for analysis in the notebook. This analysis can include subsequent visualization using standard charts provided by commonly used visualization libraries.

The categorical view supports \partialTask{filter\_jobs} by filtering the dataset on hover. The \textit{Main Summary View} and histograms update automatically to reflect filters applied this way. A user can ``pin'' this filtered view by clicking on the hovered bar. They can also add different categories to the filter by hovering over other bars and/or pinning them as well. Any affected selections will automatically update to reflect the changes made by the filter, including the records object. Already exported record objects will automatically update to reflect the change in filtered records.

Finally, hovering over a column in the \textit{Main Summary View} will update the rightmost histogram to show only the per-row records associated with that column. This linked interaction enables users to view the distribution of records for a column of interest.

\section{Evaluating Guidepost}
\label{sec:eval}
Here we describe a user study evaluating the effectiveness of our visualization for different groups working with the same data.

    \begin{table}[t!]
        \centering
     \caption{The participants (denoted P$_{\textrm{id, lab}}$) in our evaluation, their evaluation grouping, and more specific research focus. }
    \begin{tabularx}{3.5in}{l|l|X}
     & Evaluation Group & Specific Research Focus                  \\ \hline
    P$_{1A}$          & \texttt{Research Analyst}       & Jobs Data Analyst/ ML Engineer                      \\
    P$_{2A}$          & \texttt{Supercomputer User} & Domain Scientist HPC User/ Software Optimization Support   \\
    P$_{3A}$          & \texttt{Supercomputer User} & Domain Scientist HPC User                                  \\
    P$_{4A}$          & \texttt{Supercomputer User} & Domain Scientist HPC User                                  \\
    P$_{5A}$          & \texttt{Research Analyst}       & Jobs Data Analyst                                   \\
    P$_{6B}$          & \texttt{Supercomputer User} & Workflow Performance Analysis and Tooling                  \\
    P$_{7B}$          & \texttt{Research Analyst}       & Software Performance Analysis and Tooling/ Energy Research \\
    P$_{8C}$          & \texttt{Research Analyst}       & Jobs Data Analyst                                   \\
    P$_{9C}$          & \texttt{Research Analyst}       & Jobs Data Analyst/ ML Engineer
    \end{tabularx}
    \vspace{-2em}
    \label{tab:participants}
\end{table}

\subsection{Participants}

Due to the difficulty of obtaining equal representative samples for each persona, we divided our participants into two groups: a "\texttt{Research Analyst}" group and a "\texttt{Supercomputer User}" group. We capture the \ml{} and \jda{} personas in the "\texttt{Research Analyst}" group, as they both represent individuals who work primarily with HPC jobs data to understand the operations of an HPC system, even though their goals are otherwise divergent.  The "\texttt{Supercomputer User}" group represents individuals who interact with HPC jobs data as supercomputer users. They are embodied solely by the \hu{} persona.

Observing that this software would likely be helpful to many individuals working with HPC jobs data at similar organizations, we elicited feedback from potential users working at different institutions. We recruited 9 expert users from 3 US-based Federally Funded Research and Development Centers (FFRDCs) that have some need to understand HPC jobs data. Two of our participants (P$_{2A}$ and P$_{5A}$) returned from the feedback interviews we conducted on the prototype visualization. Table~\ref{tab:participants} shows our users, their laboratory affiliation, evaluation grouping, and more specific research focus.

\subsection{Study Design}
This study was conducted in 60-minute sessions either in person or via video-conferencing software by participant availability.
Each session consisted of an overview of \guidepost{} (20 minutes), two demographic questions (5 minutes), 4 tasks for the users to complete (15-20 minutes), a semi-structured interview (10 minutes), and a debriefing.

We gave participants the option to either use \textit{Guidepost} themselves or engage in pair analytics \cite{arias-hernandez_pair_2011}.
In the latter case, the facilitator acts as the ``driver'' of the visualization and the participant as the analyst telling the facilitator how to interact with the system. We offered this option based on our experience with similar notebook-based evaluations where we observed that some participants may feel self-conscious programming in front of others, inhibiting analysis. All but one participant (P$_{8C}$) chose to drive the visualization themselves.

The design of \guidepost{} differed slightly at the time this evaluation was executed from the designs shown in~\autoref{fig:full_system} and ~\autoref{fig:teaser}. We include images of the version used in the evaluations in the supplemental materials.

\inlinesec{Study Datasets:} We used a historical jobs dataset (30k records) collected across three queues serving a single FFRDC supercomputer.

This data was collected over 6 months between June and December of 2023. Each record contains predictive data produced from an in-development prediction model built by a \ml{}.

\inlinesec{Study Tasks:}

We started users with an open-ended exploration task to allow for unexpected interactions or unique insights.
We prepared them for this task by first asking if they were curious about any particular aspects of HPC jobs data. After exploration, participants were asked to perform more specific experiment tasks aligned with those analytical tasks our visualization was designed to support. We list the tasks below and indicate which analysis task(s) they evaluate in parentheses:
\vspace{-0.25em}
 \begingroup
    \renewcommand{\labelenumi}{E\theenumi}
\begin{enumerate}[left=0.25em]
    \itemsep=-0.5em
    \item Can you identify which queue among these three would likely result in the shortest wait time? (\sharedTask{compare\_q}, \sharedTask{relate\_jobs})
    \item Do you see any correlation between the number of nodes requested and wait time? (\sharedTask{assess\_q\_wait})
    \item What are the high and low points of system usage during the year (in terms of jobs submitted)? (\sharedTask{data\_context})
    \item What is the best day of the week to run an HPC job on a particular queue? (\sharedTask{filter\_jobs})
\end{enumerate}
\endgroup

These tasks were focused on answering a particular question instead of asking people to execute an interaction, simulating intended real usage. We allowed users to export and analyze data programmatically if that felt more natural.

\subsection{Evaluation/Task Results}

\inlinesec{E1 - Identify Short Queue Times:} Participants were asked to identify which queue would result in the shortest wait times overall. All participants, save one (P$_{6B}$), gave an appropriately justified answer to this question. P$_{6B}$ got involved in doing deeper analysis using exported data and the experiment facilitator moved on to the next question without probing for a focused response.

Most participants (P$_{1A}$-P$_{5A}$, P$_{7B}$, P$_{9C}$) used the vertical histogram to compare wait times between queues. In the default configuration this histogram shows distributions of queue times for each queue. Participants estimated the average wait times from this distribution and came to the conclusion that the ``short'' queue results in lower wait times overall. Besides P$_{6B}$, who began examining the data in pandas, P$_{8C}$ was the only other person who took a different approach. They made their determination using the \textit{Main Summary View} to compare over time based on which queue had fewer jobs week over week.

\inlinesec{E2 - Identify Correlations Between Variables on Dataset: } Participants were asked to identify if they observed any correlation between the number of nodes requested by an HPC job and its wait time. All participants successfully executed this task. The majority re-configured the plots (P$_{1A}$, P$_{3A}$-P$_{9C}$) to color the \textit{Main Summary View} by the number of nodes requested, keeping the y-axis on queue times and x-axis on job submit date.

With a perfect positive correlation, rectangles should be darker at the top of the summary view and lighter at the bottom. The real data shows dark squares scattered throughout a sea of mostly lighter squares. Many participants (P$_{1A}$, P$_{3A}$-P$_{7B}$, P$_{9C}$) verbally identified, despite this messiness, that darker rectangles tended to occur near the top of the plot and lighter rectangles more commonly near the bottom. P$_{6B}$ called this a ``loose correlation'' and P$_{5A}$ explained that this visualization was consistent with their understanding of the system based on other data. P$_{2A}$ flipped the y-axis and color encoding and answered this question with ``number of nodes'' as the y-axis value and the ``wait time'' aggregated into colors.

\inlinesec{E3 - Identifying high and low points of system usage: } Participants were asked to identify what points in the year this supercomputer was being used least and most. All participants successfully executed this task and answered some variation of ``late summer'' for high points and ``early fall'' for low. The approaches to this question were much more diverse than prior questions due to domain-specific nuance with the word ``utilization'' or ``usage.''  Often, system usage is measured in terms of compute hours and not just the number of jobs submitted. Some participants (P$_{1A}$, P$_{5A}$, P$_{8C}$) attempted to answer the compute hour question. In these cases, the question was refined by the facilitator to specify ``Usage, in terms of jobs submitted.'' Other participants (P$_{2A}$-P$_{4A}$, P$_{6B}$-P$_{7B}$, P$_{9C}$) identified that the bottom histogram in each row shows the number of jobs submitted per-day and can tell us when there were periods of high activity.

P$_{1A}$ and P$_{5A}$ exported selections of the data and manipulated it using pandas. As this analysis is more involved and open-ended, neither participant completed it and moved on to the simpler ``job counting'' question clarified by the facilitator. P$_{8C}$ used the visualization but focused their analysis on the temporal summary view as with E1. P$_{8C}$ used ``number of nodes requested'' (the color metric) as a surrogate metric for ``compute'' and identified days where the rectangles were darker to describe high utilization.

\inlinesec{E4 - Best day of the week to submit HPC jobs: } Participants were asked to identify what would be the best day of the week to submit a job to reduce their wait time on the standard queue. All participants completed this task with the use of a provided categorical variable, \texttt{day\_of\_week}. Participants were uniform in their approach to this task: successfully changing the visualization configuration from a default categorical variable \texttt{user} to \texttt{day\_of\_week}.

Once the categorical view in the lower-right corner showed the day of the week, participants compared the record counts between days. Many (P$_{2A}$-P$_{3A}$, P$_{6B}$-P$_{8C}$) concluded that the weekend (Saturday or Sunday) would be the best day to submit based on the job counts alone. Others (P$_{1A}$, P$_{4A}$-P$_{5A}$, P$_{9C}$) wanted to verify this intuition by filtering and comparing wait time distributions in the right histogram.

\inlinesec{Results Summary:} Participants were successful in completing tasks and did so in a variety of ways. They used all four views of each unit visualization to inspect various features on the data and also used filtering and brushing interactions. Many participants also used provided data export interactions to drop out of the visualization and into the notebook to interact with their data using familiar scripting methodologies.

\subsection{Analysis}
We discuss findings from the analysis of the semi-structured interview at the end of the evaluation and observations of user behavior.

\inlinesec{The visualization was effective for generating insights.} We observed some members of both the \texttt{Research Analyst} (P$_{1A}$, P$_{5A}$, P$_{9C}$) and \texttt{Supercomputer User} (P$_{3A}$, P$_{4A}$, P$_{6B}$) groups extracting insights about the data during the evaluation.  P$_{1A}$ identified expected supercomputer behavior based on patterns on the temporal summary view. When discussing their answer for E2, P$_{5A}$ says that the weak correlation they observe between nodes requested and wait time is consistent with their expectations. P$_{9C}$ hypothesized that a spike of activity in August relates to "people working to meet the deadline" for Supercomputing Conference workshops.

P$_{3A}$'s inquiries were focused on general system usage, uttering sentiments of surprise when given \textit{Guidepost's} dense view of the data. They elaborate, ``last week of September is when I perceived it as being super clogged up\dots the main bottleneck was here around August 20th.'' P$_{4A}$ was surprised to see how power users can be ``kind of responsible for all the hot spots.'' P$_{6B}$ echoes this insight by confirming that it conforms to their prior expectations from another work: ``90\% of the issues with the GPUs are caused by 7 users.''

\inlinesec{A majority of participants exported the data and manipulated it programmatically without explicit instruction.} The decision to develop a notebook-embedded visualization with fluid data exporting functionality is supported by our participants' actions during the evaluation. Six participants representing both the \texttt{Research Analyst} (P$_{1A}$, P$_{5A}$, P$_{8C}$, P$_{9C}$) and \texttt{Supercomputer User} groups (P$_{4A}$, P$_{6B}$) exported and examined a a sub-selection of data programmatically. For many, it was during the initial exploratory phase as they wanted to get more information about a specific feature in the data. Others used this feature to answer the more specific evaluation tasks. For some, this analysis involved looking at the content of the dataframe in another cell (P$_{5A}$, P$_{8C}$, P$_{9C}$) and going over it row-by-row. Others (P$_{1A}$, P$_{4A}$, P$_{6B}$) performed some aggregations on the dataframe to get more focused, summarized perspectives on the data.

Sentiments around this feature were very positive. In the midst of moving from the visualization context back to a scripting cell P$_{1A}$ said, ``This is awesome.''  P$_{4A}$ says that they think ``this [feature] is useful'' and that scripting is ``immediately where my head goes.'' P$_{8C}$ called this feature ``very useful.'' P$_{5A}$ said of the feature, ``To just pull out a specific subset of something you you find interesting, I think like those two things are really awesome \ldots doing this yourself manually \ldots takes a lot longer than just drag and clicking.''

\inlinesec{Constructive feedback from participants was diverse, but no weakness was a ``deal breaker.''} Feedback on pain points or areas of improvement were often specific to each individual or shared between a few individuals. They did not detract from participants' overall appreciation of the software, however. P$_{1A}$ and P$_{2A}$ both requested drop-down menus in the visualization itself to save on the effort required to apply configurations programmatically. P$_{1A}$ and P$_{9C}$ shared a desire to select export data representing one day's jobs. P$_{1A}$, P$_{2A}$, and P$_{6B}$ suggested that the scales on various histograms (including the categorical view) should support log or linear scale as appropriate to the data (at the time of evaluation, all axes were log scale only).

P$_{3A}$ and P$_{7B}$ requested more customization in the provided color map and support for a divergent color map.  P$_{7B}$ and P$_{9C}$ expressed that axes scales should be shared between the histograms in each small multiple queue. Some more notable specific features requested were support for different levels of x-axis zoom (P$_{6B}$) and support for alternative sorting of categories in the categorical view (P$_{7B}$, P$_{8C}$).

\inlinesec{Participants expressed interest in using \guidepost{} to achieve many different analysis goals.} We asked participants to describe if they would want to use this software in their notebooks or workflows. Among the \texttt{Research Analyst} group of participants (P$_{1A}$, P$_{5A}$, P$_{7B}$-P$_{9C}$), reports of intended usage aligned with our design goals. P$_{1A}$ evoked the hybrid visualization-scripting workflow we built this visualization for when elaborating on why they would use \textit{Guidepost}, saying, "I'd want to look at some aspects of the data more specifically and then I'd pull that data out just like I did." P$_{5A}$ expressed similar sentiment in different terms, "I could foresee like kind of changing my approach a little bit\ldots This 'Guidepost' first and then\ldots figure out what data I want to get and then maybe like use [the retrieve data] command to then do some subsequent things." P$_{7B}$ focused on detailing the value of the visualization itself saying it provides "a really nice first look at all of the data." P$_{8C}$ expressed that they will work to get it running with their own supercomputer's data. P$_{9C}$ asserted that they would "definitely" use a visualization like this.

\inlinesec{Limitations and Threats to Validity}
The generalizability of our evaluation is limited by the modest number of participants. We sought representative users—analysts of HPC systems data and/or users of supercomputers. These are busy professionals and are, therefore, difficult to recruit, limiting our participant pool. One hour is a typical schedule slot duration for our professional users, which limited the kinds of tasks we could include. All participants knew at least one of the authors, which may have biased their comments positively. However, many participants only had either very new relationships (P$_{8C}$, P$_{9C}$) and/or not closely tied to any individual in the research group (P$_{2A}$, P$_{3A}$, P$_{4A}$).

\section{Discussion}
\label{sec:reflect}

In this work, we investigated how to develop a single visualization that can support multiple user groups.
To support this, we conducted a design study explicitly centered on personas as a way to parse that heterogeneity in the context of HPC jobs data.
This yielded a notebook-embedded overview visualization: \textit{Guidepost}.

Via a user study with nine experts from three US-based national research laboratories, we observed that our design served the intended tasks and gave the experts tools to move easily from the visualization into programmatic workflows,  extending the utility of the tool past the tasks it directly supports.

To conclude, we reflect on incorporating personas into design projects, how to expand the audiences tools support, how users drop off of our visualizations by design, and offer considerations of how to support users where their workflows are.

\inlinesec{The use of personas for a visualization design project is not a one-size-fits-all solution but introduces flexibility that conforms well to the human-centered aspect of visualization.}
Design studies yield a wealth of qualitative data, often dramatically pruned to support discrete task-based analyses.

Personas can help re-introduce some of that qualitative richness.
A persona-based methodology's emphasis on the goals and desires of users allows designers to make broader decisions about how to fulfill these goals.

This approach does have pitfalls, however. With personas alone, design decisions can be harder to justify compared to task-based analyses which have significant support in the literature \cite{brehmer_multi-level_2013, meyer_criteria_2020}. This can be mitigated through iterative user feedback, but analyzing and understanding this feedback requires a significant time commitment.
We explicitly included prototype development into our process for this reason.

Furthermore, personas have limited utility when designing for a well-defined homogeneous user group. In those cases, a task analysis will likely capture everything a visualization designer needs to know to make a useful visualization.
Finally, the same narrative aspects that can build empathy with user's needs\cite{miaskiewicz_personas_2011}, can also lead to misaligned projection of user needs if that persona is not appropriately formed. While tasks do not risk this issue, their narrowness may miss important higher-level richness.

In this work, we developed for overlapping needs to broaden a user base. This was aided by the fact that we focused on an ``overview'' design. Not every visualization can be exploratory overviews. The needs of experts doing rigorous analysis on narrow parts of a dataset cannot be entirely forgotten to produce generally usable visualizations. Thus, we acknowledge that our methodology is not a universal solution but may be an effective part of a larger pipeline of visualizations that serve different needs at different points in an analyst's workflow.

\inlinesec{Our design process attempted to provide the most value for the most people through focused intervention.} While developing bespoke tools for each individual user would be desirable to those users, doing so is broadly not feasible due to fixed resources (such as that of the design team) and downstream maintenance difficulties \cite{akbaba_troubling_2023}. Our approach to designing for many individuals and offloading responsibilities of a visualization to other existing tools makes a targeted intervention in a manner that supports as wide a variety of users as possible.
Something that was not considered in this design strategy was the quantity of each persona---if a lab consists of 100 \ml{}s and 1 \hu{}, then a better strategy might be to center ML-focused tasks.
Our strategy was appropriate for the population we were working with, but future work might explore other approaches.

Abandonment is an essential risk for the products of many design studies, particularly those predicated on bespoke visualization tools. Some tools are designed to be used by a small number of users, and so each of those individuals offers a certain level of risk to this system (if usage is tied to monetary support). However, if we expand a user-base by deliberately designing for more individuals, the resulting software may provide unique opportunities for continued use.

\inlinesec{Additional user groups can be identified and recruited through many channels.} Although other groups may come to a project incidentally through a grapevine of interested researchers, sometimes new groups must be deliberately identified and recruited. There are many ways to identify and recruit additional users. For \guidepost{}, we were not given immediate exposure to the group that would be cast as \hu, however, we made a deliberate effort to ask system administrators who set up the original job measurement software for their data users. In a large organization, data producers can have a broad perspective on their users, which may exceed that of the initial group that sought visualization help. We also benefited from presenting preliminary results in a poster session where several potential users from unsupported groups expressed interest in our prototype for their research. These interactions can help expand a software's user base but must be handled carefully to ensure that some shared interests exist between new and existing users. Continuous use of personas can help by providing the tools to assess if new users are from a new group or would fit into an existing group.

\inlinesec{Expanding user groups may help mitigate unavoidable winnowing pitfalls in design studies.}

While winnowing is a key step in DSM, there are structural reasons why winnowing may not be possible \cite{akbaba_troubling_2023}. In cases where, for example, the origin of tasks may be initially unclear but development and design is expected due to job requirements, growing the user base can also serve the benefit of giving alternative options to save a project that may be forced into a pitfall state due to powers outside a designer's control.  A designer need not immediately discard tasks from an individual not in their target audience if they are careful when tracking the origin of said task and acting on it thoughtfully. This approach also suggests that there may be room for alternative solutions for some of the pitfalls enumerated in the Precondition phase of Design Study Methodology \cite{sedlmair_design_2012}, which the visualization community could further explore.

\inlinesec{A visualization designed with multiple users in mind should fail gracefully, by design.}

As our visualization does not support all tasks that users are interested in, we intentionally included an ``offramp'' for users. They can ``take their analysis and go'' through our support for brushing interactions and an export function. This can be conceptualized as a planned failure point. When we know that our visualization will eventually no longer support our users, providing a contingency to continue their analysis elsewhere can be helpful.

We suggest that this design philosophy empowers users with a deeper understanding of their data and equips them with selections of interesting records to visualize through another chart or ``script.'' Indicating that users may not see this as a pain point, P4 expressed this concept when discussing his impressions of \textit{Guidepost}:
\vspace{-1ex}
\begin{quote}
    ``Maybe, like the notebook or the visualization doesn't need to support multiple categories because as soon as you realize that you wanted to get all of the Mondays and then find out what users were active that day, then it becomes a scripting type of thing. You've gone as far as you can with the visualization. \ldots And then you would turn it over to some little bit of logic that you could write down here in this next cell.''
\end{quote}
\vspace{-1ex}

\guidepost{}'s offramp is a simple brush-based query that returns sub-selected data agnostically. Considering user's needs via personas offers opportunities to creatively  bridge the gap between visualization and notebook without the need for a formal task analysis and significant development effort.
For example, when a designer knows that an \ml{} is asking questions about the statistical characteristics of their data, a bespoke machine learning summary/reporting method could be introduced that outputs sub-selected data as a table of various statistical tests. Alternatively, the visualization could return a dataframe with only the features that a \ml{} is interested in. Such functionality could be written with much less effort than redesigning the layout of the visualization to incorporate a summary window which has utility to only one subgroup of users and requires additional screen real estate to manage the view.

\inlinesec{Familiarity with notebook-style scripting among users was important to the success of our evaluation.} Although not universal, we observed that many participants were particularly sympathetic to \guidepost{}'s ``visualize-first and extract-data-for-deeper-analysis'' workflow.
Our initial hypothesis, per prior work~\cite{scully-allison_design_2024}, was that less than half of the participants would engage with the export feature to look at the data. However, we observed six participants engaged with data directly, noting that it felt intuitive.
This outcome validates our choice of design but is also reflective of our user base. The individuals working with HPC jobs data (our \jda{} and \ml{} personas) are frequent users of computational notebooks. Three of four \hu{}s are domain scientists who regularly analyze scientific simulation data in computational notebooks.

This approach should be weighed carefully in future design studies. Consider where the users spend most of their time and attempt to meet them there. For some, that is a notebook; for others, it can be the command line or RStudio. In these cases, the philosophy of what we propose remains the same: leverage the tools provided by the wider environment and avoid over-designing visualization solutions.

\section{Supplemental Materials} \label{app:cc}
     The supplemental materials for this work include a \texttt{supplemental document},  a \texttt{demo video}, and a \texttt{OSF Project Repository}.

    \vspace{1em}
    \noindent The \texttt{supplemental document} contains:
    \begin{itemize}
        \itemsep=0ex
        \item Additional details and figures of the evaluation we ran in \ref{sec:eval}.
        \item Additional details and figures detailing our initial prototype for \guidepost{}.
        \item Design documents and drawings from the development process of \guidepost{}.
        \item Additional details about how personas evolved over the course of this project.
    \end{itemize}

    \vspace{1em}
    \noindent The \texttt{demo video} shows a walkthrough of \guidepost{} that explores use cases of the system from the perspective of multiple personas.

    \vspace{1em}
    \noindent The \texttt{OSF Project Repository } hosts all other documents related to the development and evaluation of \guidepost{}: \url{https://osf.io/ckbq2/?view_only=c09363fd8b0a42fc9afe0f780817231c}

    Including:
    \begin{itemize}
        \itemsep=0ex
        \item Notes collected by the first author over the course of this research.
        \item Fully anonymized transcripts and notes from the evaluation session and prototype feedback sessions.
        \item The qualitative data analysis on data collected from the evaluation and prototype feedback sessions.
        \item The source code for \guidepost{}.
        \item A downloadable jupyter notebook demonstrating \guidepost{}'s use.
    \end{itemize}

\bibliographystyle{abbrv-doi-hyperref}

\bibliography{template}

\end{document}